\documentclass{article}
\usepackage{latex8}
\usepackage{pxfonts}
\usepackage{times}
\usepackage{bbm}
\usepackage{graphicx}
\usepackage{algorithm}
\usepackage{algorithmic}
\usepackage{clrscode}
\usepackage{fullpage}
\usepackage{amssymb}
 
\begin{document}
\pagestyle{empty}

\title{A Deterministic Sub-linear Time Sparse Fourier Algorithm via Non-adaptive Compressed Sensing Methods}
\author{M. A. Iwen\thanks{Supported in part by NSF DMS-0510203.}\\
University of Michigan\\
markiwen@umich.edu}

\maketitle
\thispagestyle{empty}

\begin{abstract}
We study the problem of estimating the best $B$ term Fourier representation for a given frequency-sparse signal (i.e., vector) $\textbf{A}$ of length $N \gg B$.  More explicitly, we investigate how to deterministically identify $B$ of the largest magnitude frequencies of $\hat{\textbf{A}}$, and estimate their coefficients, in polynomial$(B,\log N)$ time.  Randomized sub-linear time algorithms which have a small (controllable) probability of failure for each processed signal exist for solving this problem.  However, for failure intolerant applications such as those involving mission-critical hardware designed to process many signals over a long lifetime, deterministic algorithms with no probability of failure are highly desirable.  In this paper we build on the deterministic Compressed Sensing results of Cormode and Muthukrishnan (CM) \cite{CMDetCS3,CMDetCS1,CMDetCS2} in order to develop the first known deterministic sub-linear time sparse Fourier Transform algorithm suitable for failure intolerant applications.  Furthermore, in the process of developing our new Fourier algorithm, we present a simplified deterministic Compressed Sensing algorithm which improves on CM's algebraic compressibility results while simultaneously maintaining their results concerning exponential decay.
\end{abstract}

\section{Introduction}

In many applications only the top few most energetic terms of a signal's Fourier Transform (FT) are of interest.  In such applications the Fast Fourier Transform (FFT), which computes all FT terms, is computationally wasteful.  To make our point, we next consider a simple application-based example in which the FFT can be replaced by faster approximate Fourier methods:\\

\noindent \textbf{Motivating Example:  sub-Nyquist frequency acquisition}

Imagine a signal/function $f: [0,2 \pi] \rightarrow \mathbb{C}$ of the form
$$f(x) = C \cdot e^{\mathbbm{i} \omega x}$$
consisting of a single unknown frequency $\omega \in (-N,N]$ (e.g., consider a windowed sinusoidal portion of a wideband frequency-hopping signal \cite{SigApp1}).  Sampling at the Nyquist-rate would dictate the need for at least $2 N$ equally spaced samples from $f$ in order to discover $\omega$ via the FFT without aliasing \cite{BoydAl}.  Thus, we would have to compute the FFT of the $2N$-length vector
$$\textbf{A}(j) = f \left( \frac{\pi j}{N} \right),~~0 \leq j < 2N.$$
However, if we use aliasing to our advantage we can correctly determine $\omega$ with significantly fewer $f$-samples taken in parallel.

Consider, for example, the two-sample Discrete Fourier Transform of $f$.  It has
$$\hat{f}(0) = C \cdot \frac{1 + (-1)^{\omega}}{2} \textrm{ and } \hat{f}(1) = C \cdot \frac{1 + (-1)^{\omega - 1}}{2}.$$ 
Clearly $\hat{f}(0) = 0$ implies that $\omega \equiv 1$ modulo 2 while $\hat{f}(1) = 0$ implies that $\omega \equiv 0$ modulo 2.  In this fashion we may use several potentially aliased Fast Fourier Transforms in parallel to discover $\omega$ modulo $3, 5, \dots,$ the $O(\log N)^{\rm th}$ prime.  Once we have collected these moduli we can reconstruct $\omega$ via the famous \textbf{Chinese Remainder Theorem (CRT)}.  

\newtheorem{Theorem}{Theorem}
\begin{Theorem}
{\sc Chinese Remainder Theorem (CRT):}  Any integer $x$ is uniquely specified mod $N$ by its remainders modulo $m$ relatively prime integers $p_{1}, \dots, p_{m}$ as long as $\prod_{l = 1}^{m} p_{l} \geq N$.
\end{Theorem}

To finish our example, suppose that $N = 500,000$ and that we have used three FFT's with 100, 101, and 103 samples to determine that $\omega \equiv 34 $ mod $100$, $\omega \equiv 3$ mod $101$, and $\omega \equiv 1$ mod 103, respectively.  Using that 
$\omega \equiv 1$ mod 103 and we can see that $\omega = 103 \cdot a + 1$ for some integer $a$.  Thus, 
$$(103 \cdot a + 1) \equiv 3 \textrm{ mod } 101 \Rightarrow 2 a \equiv 2 \textrm{ mod } 101 \Rightarrow a \equiv 1 \textrm{ mod } 101.$$
Therefore, $a = 101 \cdot b + 1$ for some integer $b$.  Substituting for $a$ we get that $\omega = 10403 \cdot b + 104$.  By similar work we can see that $b \equiv 10$ mod $100$ after considering $\omega$ modulo 100.  Hence, $\omega = 104,134$ by the CRT.  As an added bonus we note that our three FFTs will have also provided us with three different estimates of $\omega$'s coefficient $C$.

The end result is that we have used significantly less than $2N$ samples to determine $\omega$.  Using the CRT we required only $100 + 101 + 103 = 304$ samples from $f$ to determine $\omega$ since $100 \cdot 101 \cdot 103 > 1,000,000$.  In contrast, a million $f$-samples would be gathered during Nyquist-rate sampling.  Besides needing significantly less samples than the FFT, this CRT-based single frequency method dramatically reduces required computational effort.  And, it's deterministic.  There is no chance of failure.  Of course, a single frequency signal is incredibly simple.  Signals involving more than 1 non-zero frequency are much more difficult to handle since frequency moduli may begin to collide modulo various numbers.  Dealing with the potential difficulties caused by such frequency collisions in a deterministic way comprises the majority of this paper.

\subsection{Compressed Sensing and Related Work}

Compressed Sensing (CS) methods \cite{CS1,CS2,CMDetCS3,CMDetCS1,CMDetCS2} provide a robust framework for reducing the number of measurements required to estimate a sparse signal.  For this reason CS methods are useful in areas such as MR imaging \cite{MRI1, MRI2} and analog-to-digital conversion \cite{SigApp1,SigApp2} where measurement costs are high.  The general CS setup is as follows:  Let \textbf{A} be an $N$-length signal/vector with complex valued entries and $\Psi$ be a full rank $N \times N$ change of basis matrix.  Furthermore, suppose that $\Psi \cdot \textbf{A}$ is sparse (i.e., only $k \ll N$ entries of $\Psi \cdot \textbf{A}$ are significant/large in magnitude).  CS methods deal with generating a $K \times N$ measurement matrix, $\mathcal{M}$, with the smallest number of rows possible (i.e., $K$ minimized) so that the $k$ significant entries of $\Psi \cdot \textbf{A}$ can be well approximated by the $K$-element vector result of
\begin{equation}
\mathcal{M} \cdot \Psi \cdot \textbf{A}.
\label{eqn:matprod}
\end{equation}
Note that CS is inherently algorithmic since a procedure for recovering $\Psi \cdot \textbf{A}$'s largest $k$-entries from the result of Equation~\ref{eqn:matprod} must be specified.  

For the remainder of this paper we will consider the special CS case where $\Psi$ is the $N \times N$ Discrete Fourier Transform matrix.  Hence, we have
\begin{equation}
\Psi_{i,j} = e^{\frac{2 \pi \mathbbm{i} \cdot i \cdot j}{N}}
\end{equation}
Our problem of interest is to find, and estimate the coefficients of, the $k$ significant entries (i.e., frequencies) of $\hat{\textbf{A}}$ given a frequency-sparse (i.e., smooth) signal \textbf{A}.  In this case the deterministic Fourier CS measurement matrixes, $\mathcal{M} \cdot \Psi$, produced by \cite{CS2,CMDetCS3,CMDetCS1,CMDetCS2} require super-linear $O(K N)$-time to multiply by \textbf{A} in Equation~\ref{eqn:matprod}.  Similarly, the energetic frequency recovery procedure of \cite{CS1,DetCS} requires super-linear time in $N$.  Hence, none of \cite{CS1,CS2,DetCS,CMDetCS3,CMDetCS1,CMDetCS2} have both sub-linear measurement and reconstruction time.  

Existing randomized sub-linear time Fourier algorithms \cite{AAFFT1,AAFFT1exp,AAFFT2} not only show great promise for decreasing measurement costs, but also for speeding up the numerical solution of computationally challenging multi-scale problems \cite{SparseSpect, SparseSpectM}.  However, these algorithms are not deterministic and so can produce incorrect results with some small probability on each input signal.  Thus, they aren't appropriate for long-lived failure intolerant applications.

In this paper we build on the deterministic Compressed Sensing methods of Cormode and Muthukrishnan (CM) \cite{CMDetCS3,CMDetCS1,CMDetCS2} in order to construct the first known deterministic sub-linear time sparse Fourier algorithm.  In order to produce our new Fourier algorithm we must modify CM's work in two ways.  First, we alter CM's measurement construction in order to allow sub-linear time computation of Fourier measurements via aliasing.  Thus, our algorithm can deterministically approximate the result of Equation~\ref{eqn:matprod} in time $K^{2} \cdot $polylog($N$).  Second, CM use a $k$-strongly selective collection of sets \cite{Constructions} to construct their measurements for algebraically compressible signals.  We introduce the generalized notion of a $K$-majority $k$-strongly selective collection of sets which leads us to a new reconstruction algorithm with better algebraic compressibility results than CM's algorithm.  As a result, our deterministic sub-linear time Fourier algorithm has better then previously possible algebraic compressibility behavior.

The main contributions of this paper are:
\begin{enumerate}
\item We present a new deterministic compressed sensing algorithm that both $(i)$ improves on CM's algebraically compressible signal results, and $(ii)$ has comparable measurement/run time requirements to CM's algorithm for exponentially decaying signals.
\item We present the first known deterministic sub-linear time sparse DFT.  In the process, we explicitly demonstrate the connection between compressed sensing and sub-linear time Fourier transform methods.
\item We introduce $K$-majority $k$-strongly selective collections of sets which have potential applications to streaming algorithms along the lines of \cite{FirstDetCS, Crprecis}.
\end{enumerate}

The remainder of this paper is organized as follows:  In section~\ref{sec:prelim} we introduce relevant definitions and terminology.  Then, in section~\ref{sec:Measurements} we define $K$-majority $k$-strongly selective collections of sets and use them to construct our compressed sensing measurements.  Section~\ref{sec:SigReconstruct} contains our new deterministic compressed sensing algorithm along with analysis of it's accuracy and run time.  Finally, we present our deterministic sub-linear time Fourier algorithm in sections~\ref{sec:fmeasure} and~\ref{sec:inaccessible}.  Section~\ref{sec:conc} contains a short conclusion.

\section{Preliminaries}
\label{sec:prelim}

Throughout the remainder of this paper we will be interested in complex-valued functions $f: [0,2\pi] \rightarrow \mathbb{C}$ and signals (or arrays) of length $N$ containing $f$ values at various $x \in [0,2\pi]$.  We shall denote such signals by $\textbf{A}$, where $\textbf{A}(j) \in \mathbb{C}$ is the signal's $j^{th}$ complex value for all $j \in [0,N-1] \subset \mathbb{N}$.  Hereafter we will refer to the process of either calculating, measuring, or retrieving the $f$ value associated any $\textbf{A}(j) \in \mathbb{C}$ from machine memory as \textit{sampling} from $f$ and/or $\textbf{A}$.  Given a signal $\textbf{A}$ we define 
its discrete $L^{2}$-norm, or Euclidean norm, to be 
$$\| \textbf{A} \|_{2} = \sqrt{\sum_{j=0}^{N-1} |\textbf{A}(j)|^{2}}.$$  
We will also refer to $\| \textbf{A} \|_{2}^{2}$ as $\textbf{A}$'s energy.

For any signal, $\textbf{A}$, its Discrete Fourier Transform (DFT), denoted $\hat{\textbf{A}}$, is another signal of length $N$ defined as follows: 
$$\hat{\textbf{A}}(\omega) = \frac{1}{\sqrt{N}} \sum_{j=0}^{N-1} e^{\frac{-2 \pi i \omega j}{N}} \textbf{A}(j),~~~\forall \omega \in [0,N-1].$$
Furthermore, we may recover $\textbf{A}$ from its DFT via the Inverse Discrete Fourier Transform (IDFT) as follows:
$$\textbf{A}(j) = \widehat{~~\hat{\textbf{A}}~~}^{-1}(j) = \frac{1}{\sqrt{N}} \sum_{\omega=0}^{N-1} e^{\frac{2 \pi i \omega j}{N}} \hat{\textbf{A}}(\omega) ,~~~\forall j \in [0,N-1].$$
We will refer to any index, $\omega$, of $\hat{\textbf{A}}$ as a frequency.  Furthermore, we will refer to $\hat{\textbf{A}}(\omega)$ as frequency $\omega$'s 
coefficient for each $\omega \in [0,N-1]$.  Parseval's equality tells us that $\| \hat{\textbf{A}} \|_{2} = \| \textbf{A} \|_{2}$ for any signal.  In other words, the DFT 
preserves Euclidean norm and energy.  Note that any non-zero coefficient frequency will contribute to $\hat{\textbf{A}}$'s energy.  Hence, we will also refer to 
$| \hat{\textbf{A}}(\omega) |^{2}$ as frequency $\omega$'s energy.  If $| \hat{\textbf{A}}(\omega) |$ is relatively large we'll say that $\omega$ is energetic.

Our algorithm produces output of the form $(\omega_{1}, C_{1}), \dots, (\omega_{B}, C_{B})$ where each $(\omega_{j}, C_{j}) \in [0, N-1] \times \mathbb{C}$.  We will refer to any such set of $B < N$ tuples 
$$\left\{ (\omega_{j}, C_{j}) \in [0, N-1] \times \mathbb{C} \textrm{ s.t. } 1 \leq j \leq B \right\}$$ 
as a \textbf{sparse Fourier representation} and denote it with a superscript `s'.  Note that if we are given a sparse Fourier representation, $\hat{\textbf{R}}^{\rm s}$, we may consider $\hat{\textbf{R}}^{\rm s}$
to be a length-$N$ signal.  We simply view $\hat{\textbf{R}}^{s}$ as the $N$ length signal
$$\hat{\textbf{R}}(j) = \left\{ \begin{array}{ll} C_j & \textrm{if } (j,C_j) \in \widehat{\textbf{R}}^{\rm s} \\ 0 & {\rm otherwise} \end{array} \right.$$
for all $j \in [0,N-1]$.  Using this idea we may, for example, compute $\textbf{R}$ from $\hat{\textbf{R}}^{\rm s}$ via the IDFT.

A $B$ term/tuple sparse Fourier representation is $B$-optimal for a signal $\textbf{A}$ if it contains $B$ of the most energetic frequencies of $\hat{\textbf{A}}$ along with their coefficients.  More precisely, we'll say that a sparse Fourier representation 
$$\hat{\textbf{R}}^{\rm s} = \left\{ (\omega_{j}, C_{j}) \in [0, N-1] \times \mathbb{C} \textrm{ s.t. } 1 \leq j \leq B \right\}$$ 
is \textbf{$B$-optimal} for $\textbf{A}$ if there exists a valid ordering of $\hat{\textbf{A}}$'s coefficients by magnitude
\begin{equation}
|\hat{\textbf{A}}(\omega_{1})| \geq |\hat{\textbf{A}}(\omega_{2})| \geq \dots \geq |\hat{\textbf{A}}(\omega_{j})| \geq \dots \geq |\hat{\textbf{A}}(\omega_{N})|
\label{eqn:ordering}
\end{equation}
so that $\big\{ (\omega_{l},\hat{\textbf{A}}(\omega_{l}))~\big|~l \in [1,B] \big\} = \hat{\textbf{R}}^{\rm s}$.  Note that a signal may have several $B$-optimal Fourier 
representations if its frequency coefficient magnitudes are non-unique.  For example, there are two 1-optimal sparse Fourier representations for the signal 
$$\textbf{A}(j) = 2 e^{\frac{2 \pi i j}{N}} + 2 e^{\frac{4 \pi i j}{N}},~N > 2.$$
However, all $B$-optimal Fourier representations, $\hat{\textbf{R}}^{\rm s}_{\rm opt}$, for any signal \textbf{A} will always have both the same unique $\| \textbf{R}_{\rm opt} \|_{2}$ and $\| \textbf{A} - \textbf{R}_{\rm opt} \|_{2}$ values.  

We continue with two final definitions:  Let $\omega_{b}$ be a $b^{th}$ most energetic frequency as per Equation~\ref{eqn:ordering}.  We will say that a signal $\hat \textbf{A}$ is \textbf{(algebraically) $p$-compressible} for some $p > 1$ if $| \textbf{A}(\omega_{b}) | = O(b^{-p})$ for all $b \in [1,N)$.  If $\textbf{R}^{\rm s}_{\rm opt}$ is a $B$-optimal Fourier representation we can see that
\begin{equation}
\| \textbf{A} - \textbf{R}_{\rm opt} \|^{2}_{2} = \sum^{N-1}_{b = B} |\textbf{A}(\omega_{b})|^{2}_{2} = O \left( \int^{\infty}_{B} b^{-2p} db \right) = O( B^{1-2p} ).
\label{equ:pbound}
\end{equation}
Hence, any $p$-compressible signal $\textbf{A}$ (i.e., any signal with a fixed $c \in \mathbb{R}$ so that $| \textbf{A}(\omega_{b}) |_{2} \leq c \cdot b^{-p}$ for all $b \in [1,N)$) will have $\| \textbf{A} - \textbf{R}^{\rm opt}_{B} \|^{2}_{2} \leq \tilde{c}_{p} \cdot B^{1-2p}$ for some $\tilde{c}_{p} \in \mathbb{R}$.  For any $p$-compressible signal class (i.e., for any choice of $p$ and $c$) we will refer to the related optimal $O(B^{1-2p})$-size worst case error value (i.e., Equation~\ref{equ:pbound} above) as $\| C^{\rm opt}_{B} \|^{2}_{2}$.  Similarly, we define an \textbf{exponentially compressible} (or \textbf{exponentially decaying}) signal for a fixed $\alpha$ to be one for which
$| \hat{\textbf{A}}(\omega_{b}) | = O(2^{- \alpha b})$.  The optimal worst case error is then
\begin{equation}
\| C^{\rm opt}_{B} \|^{2}_{2} = O \left( \int^{\infty}_{B} 4^{- \alpha b} db \right) = O(4^{- \alpha B}).
\label{eqn:expbound}
\end{equation}

Fix $\delta$ small (e.g., $\delta = 0.1$).
Given a compressible input signal, $\textbf{A}$, our deterministic Fourier algorithm will identify $B$ of the most energetic frequencies from $\hat{\textbf{A}}$ and approximate their coefficients to produce a Fourier representation $\hat{\textbf{R}}^{\rm s}$ with $\| \textbf{A} - \textbf{R} \|^{2}_{2} \leq \| \textbf{A} - \textbf{R}_{\rm opt} \|^{2}_{2} + \delta \| C^{\rm opt}_{B} \|^{2}_{2}$.  These are the same types of compressible signal results proven by CM \cite{CMDetCS1,CMDetCS2}.

\section{Construction of Measurements}
\label{sec:Measurements}

We will use the following types of subset collections to form our measurements:

\newtheorem{Definition}{Definition}
\begin{Definition}
A collection, $\mathcal{S}$, of $s$ subsets of $[0,N)$ is called \textbf{$K$-majority $k$-strongly selective} if for any $X \subset [0,N)$ with $| X | \leq k$, and for all $x \in X$, the following are true:  $(i)$ $x$ belongs to $K$ subsets in $\mathcal{S}$ and, $(ii)$ more than two-thirds of $S_{j} \in \mathcal{S}$ containing $x$ are such that $S_{j} \cap X = \{ x \}$ (i.e., every member of $X$ occurs separated from all other members of $X$ in more than two-thirds of the $K$ $\mathcal{S}$-subsets it belongs to).
\label{def:SepSets}
\end{Definition}

A 1-majority $k$-strongly selective collection of sets is an example of a \textbf{$k$-strongly selective collection of sets} \cite{Constructions,CMDetCS3}.  Note that a $K$-majority $k$-strongly selective collection of subsets contains many $k$-strongly selective collections of subsets (i.e., it has repeated strong selectivity).  Thus, our newly defined $K$-majority $k$-strongly selective collections are help us count how many times each small subset element is isolated.  This added structure allows a new reconstruction algorithm (Algorithm~\ref{alg:reconstruct}) with better algebraic compressibility properties than previous methods.

Next, we will build $O(\log N)$ $K$-majority $k$-strongly selective collections of subsets.  Each of these $O(\log N)$ collections will ultimately be used to determine energetic frequencies modulo a small prime $< N$.  These moduli will then be used along with the Chinese Remainder Theorem to reconstruct each energetic frequency in a manner akin to the introduction's motivating example.  Our technique is motivated by the method of prime groupings first employed in \cite{FirstDetCS}.  To begin, we will denote each of the $O(\log N)$ collections of subsets by $\mathcal{S}_{l}$ where $0 \leq l \leq O(\log N)$.  We construct each of these $K$-majority $k$-strongly selective collections as follows:  

Define $p_{0} = 1$ and let 
$$p_{1}, p_{2}, \dots, p_{l}, \dots, p_{m}$$ 
be the first $m$ primes where $m$ is such that
$$\prod^{m-1}_{l=1} p_{l} \leq \frac{N}{k} \leq \prod^{m}_{l=1} p_{l}.$$
Hence, $p_{l}$ is the $l^{\rm th}$ prime natural number and we have
$$p_{0} = 1, p_{1} = 2, p_{2} = 3, p_{3} = 5, \dots, p_{m} = O(m \log m).$$
Note that we know $p_{m} = O(m \log m)$ via the Prime Number Theorem, and so $p_{m} = O(\log N \log\log N)$.  Each $p_{l}$ will correspond to a different $K$-majority $k$-strongly selective collection of subsets of $[0,N) = \{0, \dots, N-1 \}$.

Along the same lines we let $q_{1}$ through $q_{K}$ be the first $K$ (to be specified later) consequitive primes such that
$$\max( p_{m}, k ) \leq q_{1} \leq q_{2} \leq \dots \leq q_{K}.$$  
We are now ready to build $\mathcal{S}_{0}$, our first $K$-majority k-strongly selective collection of sets.  We begin by letting $S_{0,j,h}$ for all $1 \leq j \leq K$ and $0 \leq h \leq q_{j}-1$ be
$$S_{0,j,h} = \{ n \in [0,N)~|~n \equiv h \textrm{ mod } q_{j} \}.$$
Next, we progressively define $S_{0,j}$ to be all integer residues mod $q_{j}$, i.e., 
$$S_{0,j} = \{ S_{0,j,h}~|~ h \in [0,q_{j}) \},$$
and conclude by setting $\mathcal{S}_{0}$ equal to all $K$ such $q_j$ residue groups:
$$\mathcal{S}_{0} = \cup^{K}_{j = 1} S_{0,j}.$$
More generally, we define $\mathcal{S}_{l}$ for $0 \leq l \leq m$ as follows:
$$\mathcal{S}_{l} = \cup^{K}_{j=1} \left\{ \{ n \in [0,N)~|~n \equiv h \textrm{ mod } p_{l}q_{j} \}~\big|~h \in [0,p_{l}q_{j}) \right\}.$$

\newtheorem{Lemma}{Lemma}
\begin{Lemma}
Fix $k$.  If we set $K \geq 3 (k - 1) \lfloor \log_{k} N \rfloor + 1$ then $\mathcal{S}_{0}$ will be a $K$-majority $k$-strongly selective collection of sets.  Furthermore, if $K = O(k \log_{k} N)$ then $| \mathcal{S}_{0} | = O\left(k^{2} \log^{2}_{k} N \cdot \max(\log k, \log \log_{k} N )\right)$.
\label{lem:S0strong}
\end{Lemma}

\noindent \textit{Proof:}\\

Let $X \subset [0,N)$ be such that $|X| \leq k$.  Furthermore, let $x,y \in X$ be such that $x \neq y$.  By the Chinese Remainder Theorem we know that $x$ and $y$ may only collide modulo at most $\lfloor \log_{k} N \rfloor$ of the $K$ $q$-primes $q_{K} \geq \dots \geq q_{1} \geq k$.  Hence, $x$ may collide with all the other elements of $X$ (i.e., with $X - \{x \}$) modulo at most $(k - 1) \lfloor \log_{k} N \rfloor$ $q$-primes.  We can now see that $x$ will be isolated from all other elements of $X$ modulo at least $K - (k - 1) \lfloor \log_{k} N \rfloor \geq 2 (k - 1) \lfloor \log_{k} N \rfloor + 1 > \frac{2K}{3}$ $q$-primes.  This leads us to the conclusion that $\mathcal{S}_{0}$ is indeed $K$-majority $k$-strongly selective.

Finally, we have that
$$| \mathcal{S}_{0} | \leq \sum^{K}_{j=1} q_{j} \leq K \cdot q_{K}.$$
Furthermore, given that $K > \max(k,m)$, the Prime Number Theorem tells us that $q_{K} = O(K \log K)$.  Thus, we can see that $\mathcal{S}_{0}$ will indeed contain $O\left(k^{2} \log^{2}_{k} N \cdot \max(\log k, \log \log_{k} N )\right)$ sets.~~$\Box$ \\

Note that at least $O(k\log_{k} N)$ primes are required in order to create a ($K$-majority) $k$-strongly separating collection of subsets using primes in this fashion.  Given any $x \in [0,N)$ a $k-1$ element subset $X$ can be created via the Chinese Remainder Theorem and $x$ moduli so that every element of $X$ collides with $x$ in any desired $O(\log_{k} N)$ $q$-primes.  We next consider the properties of the other $m$ collections we have defined:  $\mathcal{S}_{1}, \dots, \mathcal{S}_{m}$.

\begin{Lemma}
Let $S_{l,j,h} = \{ n \in [0,N)~|~n \equiv h \textrm{ mod } p_{l}q_{j} \}$, $X \subset [0,N)$ have $\leq k$ elements, and $x \in X$.  Furthermore, suppose that $S_{0,j,h} \cap X = \{ x \}$. Then, for all $l \in [1,m]$, there exists a unique $b \in [0,p_{l})$ so that $S_{l,j,h + b \cdot q_{j}} \cap X = \{ x \}$.
\label{lem:S1-m}
\end{Lemma}

\noindent \textit{Proof:}\\

Fix any $l \in [1,m]$.  $S_{0,j,h} \cap X = \{ x \}$ implies that $x = h + a \cdot q_{j}$ for some unique integer $a$.  Using $a$'s unique representation modulo $p_{l}$ (i.e., $a = b + c \cdot p_{l}$) we get that $x = h + b \cdot q_{j} + c \cdot q_{j} p_{l}$.  Hence, we can see that $x \in S_{l,j,h + bq_{j}}$.  Furthermore, no other element of $X$ is in $S_{l,j,h + t \cdot q_{j}}$ for any $t \in [0,p_{l})$ since it's inclusion therein would imply that it was also an element of $S_{0,j,h}$.~~$\Box$ \\

Note that Lemma~\ref{lem:S1-m} and Lemma~\ref{lem:S0strong} together imply that each $\mathcal{S}_{1}, \dots, \mathcal{S}_{m}$ is also a $K$-majority $k$-strongly separating collection of subsets.  Also, we can see that if $x \in S_{l,j,h + b \cdot q_{j}}$ we can find $x$ mod $p_{l}$ by simply computing $h + b q_{j}$ mod $p_{l}$.  Finally, we form our measurement matrix.

Set $\mathcal{S} = \cup^{m}_{l = 0} \mathcal{S}_{l}$.  To form our measurement matrix, $\mathcal{M}$, we simply create one row for each $S_{l,j,h} \in \mathcal{S}$ by computing the $N$-length characteristic function vector of $S_{l,j,h}$, denoted $\chi_{S_{l,j,h}}$.  This leads to $\mathcal{M}$ being a $\tilde{O}(k^{2})$ x $N$ measurement matrix.  Here we bound the number of rows in $\mathcal{M}$ by noting that: $(i)$ $| \mathcal{S} | < m \cdot K \cdot p_{m}q_{K}$, $(ii)$ $m = O(\log N)$, $(iii)$ $p_{m} = O(\log N \cdot \log\log N)$, $(iv)$ $K = O( k \log N)$, and $(v)$ $q_{K} = O(K \log K)$.

\section{Signal Reconstruction from Measurements}
\label{sec:SigReconstruct}

Let $\hat{\textbf{A}}$ be an $N$-length signal of complex numbers with it's $N$ entries numbered 0 through $N-1$.  Our goal is to identify $B$ of the largest magnitude entries of $\hat\textbf{A}$ (i.e., the first $B$ entries in a valid ordering of $\hat{\textbf{A}}$ as in Equation~\ref{eqn:ordering}) and then estimate their signal values.  Toward this end, set 
\begin{equation}
\epsilon = \frac{|\hat{\textbf{A}}(\omega_{B})|}{\sqrt{2} C}
\label{eqn:epsilon}
\end{equation}
where $C > 1$ is a constant to be specified later, and let $B'$ be the smallest integer such that 
\begin{equation}
\sum^{N-1}_{b=B'} | \hat{\textbf{A}}(\omega_{b}) | < \frac{\epsilon}{2}.
\label{def:B'}
\end{equation}
Note that $B'$ is defined to be the last possible significant frequency (i.e., with energy $>$ a fraction of $|\hat{\textbf{A}}(\omega_{B})|$).
We expect to work with sparse/compressible signals so that $B \leq B' \ll N$.  Later we will give specific values for $C$ and $B'$ depending on $B$, the desired approximation error, and $\hat{\textbf{A}}$'s compressibility characteristics.  For now we show that we can identify/approximate $B$ of $\hat{\textbf{A}}$'s largest magnitude entries each to within $\epsilon$-precision via Algorithm~\ref{alg:reconstruct}.

\begin{algorithm}[tb]
\begin{algorithmic}[1]
\caption{$\proc{Sparse Approximate}$} \label{alg:reconstruct}
\STATE \textbf{Input: Signal $\hat{\textbf{A}}$, integers $B, B'$} 
\STATE \textbf{Output: $\hat{\textbf{R}}^{\rm s}$, a sparse representation for $\hat{\textbf{A}}$}
\STATE Initialize $\hat{\textbf{R}}^{\rm s} \leftarrow \emptyset$
\STATE Set $K = 3 B'\lfloor \log_{B'} N \rfloor + 1$
\STATE Form measurement matrix, $\mathcal{M}$, via $K$-majority $B'$-strongly selective collections (Section~\ref{sec:Measurements})
\STATE Compute $\mathcal{M} \cdot \hat{\textbf{A}}$ \\
\begin{center}
{\sc Identification}
\end{center}
\FOR {$j$ from $1$ to $K$}
	\STATE Sort $\langle \chi_{S_{0,j,0}},\hat{\textbf{A}} \rangle, \langle \chi_{S_{0,j,1}},\hat{\textbf{A}} \rangle, \dots, \langle \chi_{S_{0,j,q_{j}-1}},\hat{\textbf{A}} \rangle$ by magnitude
	\FOR {$b$ from $1$ to $B'+1$}
		\STATE $k_{j,b} \leftarrow b^{\rm th}$ largest magnitude $\langle \chi_{S_{0,j,\cdot}},\hat{\textbf{A}} \rangle$-measurement
		\STATE $r_{0,b} \leftarrow$ $k_{j,b}$'s associated residue mod $q_{j}$ (i.e., the $\cdot$ in $\langle \chi_{S_{0,j,\cdot}},\hat{\textbf{A}} \rangle$)
		\FOR {$l$ from $1$ to $m$}
			\STATE $t_{\rm min} \leftarrow \min_{t \in [0,p_{l})} |k_{j,b} - \langle \chi_{S_{l,j,t \cdot q_{j} + r_{0,b}}},\hat{\textbf{A}} \rangle|$
			\STATE $r_{l,b} \leftarrow r_{0,b} + t_{\rm min} \cdot q_{j}$ mod $p_{l}$
		\ENDFOR
		\STATE Construct $\omega_{j,b}$ from $r_{0,b}, \dots, r_{m,b}$ via the Chinese Remainder Theorem
	\ENDFOR
\ENDFOR \\
\STATE Sort $\omega_{j,b}$'s maintaining duplicates and set $C(\omega_{j,b}) =$ the number of times $\omega_{j,b}$ was constructed via line 16
\begin{center}
{\sc Estimation}
\end{center}
\FOR {$j$ from $1$ to $K$}
	\FOR {$b$ from $1$ to $B' + 1$}	
		\IF { $C(\omega_{j,b}) > \frac{2K}{3}$}
			\STATE $C(\omega_{j,b}) \leftarrow 0$ 
			\STATE $x = {\rm median} \{ {\rm real}(k_{j',b'}) | \omega_{j',b'} = \omega_{j,b} \}$
			\STATE $y = {\rm median} \{ {\rm imag}(k_{j',b'}) | \omega_{j',b'} = \omega_{j,b} \}$
			\STATE $\hat{\textbf{R}}^{\rm s} \leftarrow \hat{\textbf{R}}^{s} \cup \{ (\omega_{j,b}, x+ \mathbbm{i} y) \}$
		\ENDIF
	\ENDFOR
\ENDFOR \\
\STATE Output $B$ largest magnitude entries in $\hat{\textbf{R}}^{\rm s}$
\end{algorithmic}
\end{algorithm}

Algorithm~\ref{alg:reconstruct} works by using $\mathcal{S}_{0}$ measurements to separate $\hat{\textbf{A}}$'s significantly energetic frequencies $\Omega = \{ \omega_{1}, \dots, \omega_{B'} \} \subset [0,N)$.  Every measurement which successfully separates an energetic frequency $\omega_{j}$ from all other members of $\Omega$ will both $(i)$ provide a good (i.e., within $\frac{\epsilon}{2} \leq \frac{|\hat{\textbf{A}}(\omega_{B})|}{2 \sqrt{2}}$) coefficient estimate for $\omega_{j}$, and $(ii)$ yield information about $\omega_{j}$'s identity.  Frequency separation occurs because our $\mathcal{S}_{0}$ measurements can't collide any fixed $\omega_{j} \in \Omega$ with any other member of $\Omega$ modulo more then $(B' - 1) \log_{B'} N$ $q$-primes (see Lemma~\ref{lem:S0strong}).  Therefore, more than $\frac{2}{3}^{\rm rds}$ of $\mathcal{S}_{0}$'s $3B'\log_{B'} N + 1$ $q$-primes will isolate any fixed $\omega_{j} \in \Omega$.  This means that our reconstruction algorithm will identify all frequencies at least as energetic as $\omega_{B}$ at least $2B'\log_{B'} N + 1$ times.  We can ignore any frequencies that aren't recovered this often.  On the other hand, for any frequency that is identified more then $2B'\log_{B'} N$ 
times, at most $B'\log_{B'} N$ of the measurements which lead to this identification can be significantly contaminated via collisions with $\Omega$ members.  Therefore, we can take a median of the more than $2B'\log_{B'} N$ measurements leading to the recovery of each frequency as that frequency's coefficient estimate.  Since more than half of these measurements must be accurate, the median will be accurate.  The following Theorem is proved in the appendix.

\begin{Theorem}
Let $\hat{\textbf{R}}_{\rm opt}$ be a $B$-optimal Fourier representation for our input signal \textbf{A}.  Then, the $B$ term representation $\hat{\textbf{R}}^{\rm s}$ returned from Algorithm~\ref{alg:reconstruct} is such that $\| \textbf{A} - \textbf{R}\|^{2}_{2} \leq \| \textbf{A} - \textbf{R}_{\rm opt} \|^{2}_{2} + \frac{6 B \cdot |\hat{\textbf{A}}(\omega_{B})|^{2}}{C}$.  Furthermore, Algorithm~\ref{alg:reconstruct}'s Identification and Estimation (lines 7 - 30) run time is $O(B'^{2} \log^{4} N )$.  The number of measurements used is $O( B'^{2} \log^{6} N )$.
\label{thm:alg1}
\end{Theorem}

Theorem~\ref{thm:alg1} immediately indicates that Algorithm~\ref{alg:reconstruct} gives us a deterministic $O(B^{2} \log^{6} N)$-measurement, $O(B^{2} \log^{4} N)$-reconstruction time method for exactly recovering $B$-support vectors.  If $\hat{\textbf{A}}$ is a $B$-support vector then setting $B' = B + O(1)$ and $C = 1$ will be sufficient to guarantee that both $|\hat{\textbf{A}}(\omega_{B})|^{2} = 0$ and $\sum^{N-1}_{b=B'} | \hat{\textbf{A}}(\omega_{b}) | = 0$ are true.  Hence, we may apply Theorem~\ref{thm:alg1} with $B' = B + O(1)$ and $C = 1$ to obtain a perfect reconstruction via Algorithm~\ref{alg:reconstruct}.  However, we are mainly interested in the more realistic cases where $\hat{\textbf{A}}$ is either algebraically or exponentially compressible.  The following theorem (proved in the appendix) presents itself.

\begin{Theorem}
Let $\hat{\textbf{A}}$ be $p$-compressible.  Then, Algorithm~\ref{alg:reconstruct} can return a $B$ term sparse representation $\hat{\textbf{R}}^{\rm s}$ with $\| \textbf{A} - \textbf{R}\|^{2}_{2} \leq \| \textbf{A} - \textbf{R}_{\rm opt} \|^{2}_{2} + \delta \| C^{\rm opt}_{B} \|^{2}_{2}$ using $O \left( B^{\frac{2p}{p-1}} \delta^{\frac{2}{1-p}} \log^{4} N \right)$ total identification/estimation time and $O \left( B^{\frac{2p}{p-1}} \delta^{\frac{2}{1-p}} \log^{6} N \right)$ measurements.  If $\hat{\textbf{A}}$ decays exponentially, Algorithm~\ref{alg:reconstruct} can return a B term sparse representation, $\hat{\textbf{R}}^{\rm s}$, with $\| \textbf{A} - \textbf{R}\|^{2}_{2} \leq \| \textbf{A} - \textbf{R}_{\rm opt} \|^{2}_{2} + \delta \| C^{\rm opt}_{B} \|^{2}_{2}$ using both $\left( B^{2} + \log^{2} \delta^{\frac{-1}{\alpha}} \right) \cdot$ polylog($N$) measurements and identification/estimation time.
\label{thm:compress}
\end{Theorem}

For $p$-compressible signals, $p > 2$, CM's algorithm \cite{CMDetCS1,CMDetCS2} takes $O \left( B^{\frac{6p}{p-2}}\delta^{\frac{6}{2-p}} \log^{6} N \right)$- identification/estimation time and $O \left( B^{\frac{4p}{p-2}}\delta^{\frac{4}{2-p}} \log^{4} N \right)$-measurements to achieve the same error bound.  As a concrete comparison, CM's algorithm requires $O(B^{18} \delta^{-6} \log^{6} N)$- identification/estimation time and $O(B^{12} \delta^{-4} \log^{4} N)$-measurements for 3-compressible signals.  Algorithm~\ref{alg:reconstruct}, on the other hand, requires only $O(B^{3} \delta^{-1} \log^{4} N)$- identification/estimation time and $O( B^{3} \delta^{-1} \log^{6} N)$-measurements.  Hence, we have improved on CM's algebraic compressibility results.  All that's left to do in order to develop a deterministic sub-linear time Fourier algorithm is to compute our CS Fourier measurements (Algorithm~\ref{alg:reconstruct} lines 1 - 6) in sub-linear time.

\section{Sub-linear Time Fourier Measurement Acquisition}
\label{sec:fmeasure}

\begin{algorithm}[tb]
\begin{algorithmic}[1]
\caption{$\proc{Fourier Measure}$} \label{alg:fmeasure}
\STATE \textbf{Input:  $f$-samples, integers $m, K$} 
\STATE \textbf{Output:  $<\chi_{S_{l,j,h}}, \hat{f}>$-measurements}
\STATE Zero a $O(q_{K}p_{m})$-element array, \textbf{A} 
\FOR {$j$ from $1$ to $K$}
	\FOR {$l$ from $1$ to $m$}
		\STATE $\textbf{A} \leftarrow f(0), f \left(\frac{2 \pi}{q_{j}p_{l}} \right), \dots, f \left( \frac{2 \pi (q_{j}p_{l} - 1)}{q_{j}p_{l}} \right)$
		\STATE Calculate $\hat{\textbf{A}}$ via FFT
		\STATE $<\chi_{S_{l,j,h}}, \hat{f}> \leftarrow \hat{\textbf{A}}(h)$ for each $h \in [0,q_{j}p_{l})$
	\ENDFOR
\ENDFOR \\
\STATE Output $<\chi_{S_{l,j,h}}, \hat{f}>$-measurements
\end{algorithmic}
\end{algorithm}

Our goal in this section is to demonstrate how to use Algorithm~\ref{alg:reconstruct} as means to approximate the Fourier transform of a signal/function $f: [0, 2 \pi] \rightarrow \mathbb{C}$, where $(i)$ $f$ has an integrable $p^{\rm th}$ derivative, and $(ii)$ $f(0) = f(2 \pi), f'(0) = f'(2 \pi), \dots, f^{(p-2)}(0) = f^{(p-2)}(2 \pi)$.  In this case we know the Fourier coefficients for $f$ to be $p$-compressible \cite{BoydAl,FourierCont}.  Hence, for $N = q_{1} \cdot p_{1} \cdots p_{m}$ sufficiently large, if we can collect the necessary Algorithm~\ref{alg:reconstruct} (line 5 and 6) measurements in sub-linear time we will indeed be able to use Algorithm~\ref{alg:reconstruct} as a sub-linear time Fourier algorithm for $f$.  

Note that in order to validate the use of Algorithm~\ref{alg:reconstruct} (or any other sparse approximate Fourier Transform method \cite{AAFFT1,AAFFT2}) we must assume that $f$ exhibits some multi-scale behavior.  If $\hat{f}$ contains no unpredictably energetic large (relative to the number of desired Fourier coefficients) frequencies then it is more computationally efficient to simply use standard FFT/USFFT methods \cite{FFT,AUSFFTrev,AUSFFT,USFFT1,USFFT2}.  The responsible user, therefore, is not entirely released from the obligation to consider $\hat{f}$'s likely characteristics before proceeding with computations.

Choose any Section~\ref{sec:Measurements} $q$-prime $q_{j}$, $j \in [1,K]$, and any $p$-prime $p_{l}$ with $l \in [0,m]$.  Furthermore, pick $h \in [0,q_{j}p_{l})$.  Throughout the rest of this discussion we will consider $f$ to be accessible to sampling at any desired predetermined positions $t \in [0,2 \pi]$.  Given this assumption we may sample $f$ at $t = 0, \frac{2 \pi}{q_{j}p_{l}}, \dots, \frac{2 \pi (q_{j}p_{l} - 1)}{q_{j}p_{l}}$ in order to perform the following DFT computation:
$$ <\chi_{S_{l,j,h}}, \hat{f}> = \frac{1}{q_{j}p_{l}} \sum^{q_{j}p_{l}-1}_{k = 0} f \left( \frac{2 \pi k}{q_{j}p_{l}} \right) e^{\frac{- 2 \pi \mathbbm{i} h k}{q_{j}p_{l}}}.$$
Via aliasing \cite{BoydAl} this reduces to
$$\frac{1}{q_{j}p_{l}} \sum^{q_{j}p_{l}-1}_{k = 0} f \left( \frac{2 \pi k}{q_{j}p_{l}} \right) e^{\frac{- 2 \pi \mathbbm{i} h k}{q_{j}p_{l}}} = \frac{1}{q_{j}p_{l}} \sum^{q_{j}p_{l}-1}_{k = 0} \left( \sum^{\infty}_{\omega = - \infty} \hat{f}(\omega) e^{\frac{2 \pi \mathbbm{i} \omega k}{q_{j}p_{l}}} \right) e^{\frac{- 2 \pi \mathbbm{i} h k}{q_{j}p_{l}}} = \frac{1}{q_{j}p_{l}} \sum^{\infty}_{\omega = - \infty} \hat{f}(\omega) \sum^{q_{j}p_{l}-1}_{k = 0} e^{\frac{2 \pi \mathbbm{i} (\omega - h) k}{q_{j}p_{l}}} = \sum_{\omega \equiv h \textrm{ mod } q_{j}p_{l}} \hat{f}(\omega).$$
Using Sections~\ref{sec:Measurements} and \ref{sec:SigReconstruct} we can see that these measurements are exactly what we need in order to determine $B$ of the most energetic frequencies of $\hat{f}$ modulo $N = q_{1} \cdot p_{1} \cdots p_{m}$ (i.e., $B$ of the most energetic frequencies of $f$'s band-limited interpolant's DFT).  

We are now in the position to modify Algorithm~\ref{alg:reconstruct} in order to find a sparse Fourier representation for $\hat{f}$.  To do so we proceed as follows:  First, remove lines 5 and 6 and replace them with Algorithm~\ref{alg:fmeasure} for computing all the necessary $<\chi_{S_{l,j,h}}, \hat{f}>$-measurements.  Second, replace each ``$<\chi_{S_{l,j,h}}, \hat{\textbf{A}}>$" by ``$<\chi_{S_{l,j,h}}, \hat{f}>$" in Algorithm~\ref{alg:reconstruct}'s {\sc Identification} section.  It remains to show that these Algorithm~\ref{alg:reconstruct} modifications indeed yield a sub-linear time approximate Fourier transform.  The following theorem presents itself (see appendix for proof):

\begin{Theorem}
Let $f: [0,2 \pi] \rightarrow \mathbb{C}$ have $(i)$ an integrable $p^{\rm th}$ derivative, and $(ii)$ $f(0) = f(2 \pi), f'(0) = f'(2 \pi), \dots, f^{(p-2)}(0) = f^{(p-2)}(2 \pi)$ for some $p > 1$.  Furthermore, assume that $\hat{f}$'s $B' = O \left( B^{\frac{2p}{p-1}} \epsilon^{\frac{2}{1-p}} \right)$ largest magnitude frequencies all belong to $\left( - \big\lceil \frac{N}{2} \big\rceil, \big\lfloor \frac{N}{2} \big\rfloor \right]$.  Then, we may use Algorithm~\ref{alg:reconstruct} to return a $B$ term sparse Fourier representation, $\hat{\textbf{R}}^{\rm s}$, for $\hat{f}$ such that $\| \hat{f} - \hat{\textbf{R}} \|^{2}_{2} \leq \| \hat{f} - \hat{\textbf{R}}_{\rm opt} \|^{2}_{2} + \delta \| C^{\rm opt}_{B} \|^{2}_{2}$ using $O \left( B^{\frac{2p}{p-1}} \delta^{\frac{2}{1-p}} \log^{7} N \right)$-time and $O \left( B^{\frac{2p}{p-1}} \delta^{\frac{2}{1-p}} \log^{6} N \right)$-measurements from $f$.
\label{thm:subfourier}
\end{Theorem}

If $f: [0,2 \pi] \rightarrow \mathbb{C}$ is smooth (i.e., has infinitely many continuous derivatives on the unit circle where 0 is identified with $2 \pi$) it follows from Theorem~\ref{thm:subfourier} that Algorithm~\ref{alg:reconstruct} can be used to find an $\delta$-accurate, with $\delta = O \left( \frac{1}{N} \right)$, sparse $B$-term Fourier representation for $\hat{f}$ using $\tilde{O}(B^{2})$-time/measurements.  This result differs from previous sub-linear time Fourier algorithms \cite{AAFFT1,AAFFT2} in that both the algorithm and the measurements/samples it requires are deterministic.  Recall that the deterministic nature of the algorithm's required samples is potentially beneficial for failure intolerant hardware.  In signal processing applications the sub-Nyquist sampling required to compute Algorithm~\ref{alg:reconstruct}'s $<\chi_{S_{l,j,h}}, \hat{f}>$-measurements could be accomplished via $\tilde{O}(B)$ parallel low-rate analog-to-digital converters.

\subsection{DFT from Inaccessible Signal Samples}
\label{sec:inaccessible}

Throughout the remainder of this section we will consider our $N$-length compressible vector $\hat{\textbf{A}}$ to be the product of the $N$ x $N$ DFT matrix, \textbf{$\Psi$}, and a non-sparse $N$-length vector \textbf{A}.  Thus,
$$\hat{\textbf{A}} = \Psi \textbf{A}.$$
Furthermore, we will assume that \textbf{A} contains equally spaced samples from some unknown smooth function $f: [0,2 \pi] \rightarrow \mathbb{C}$ (e.g., \textbf{A}'s band-limited interpolent).  Hence,
$$\textbf{A}(j) = f \left( \frac{2 \pi j}{N} \right),~j \in [0,N).$$
We would like to use our modified Algorithm~\ref{alg:reconstruct} along with Algorithm~\ref{alg:fmeasure} to find a sparse Fourier representation for $\hat{\textbf{A}}$.  However, unless $\frac{N}{q_{j}p_{l}} \in \mathbb{N}$ for all $q_{j}p_{l}$-pairs (which would imply $f$ had been grossly oversampled), \textbf{A} won't contain all the $f$-samples required by Algorithm~\ref{alg:fmeasure}.  Not having access to $f$ directly, and restricting ourselves to sub-linear time approaches only, we have little recourse but to locally interpolate $f$ around our required samples.

For each required Algorithm~\ref{alg:fmeasure} $f$-sample at $t = \frac{2 \pi h}{q_{j}p_{l}}, h \in [0,q_{j}p_{l}),$ we may approximate $f(t)$ to within $O(N^{-2 \kappa})$-error by constructing 2 local interpolents (one real, one imaginary) around $t$ using \textbf{A}'s nearest $2 \kappa$ entries \cite{BasicInterp}.  These errors in $f$-samples can lead to errors of size $O( N^{-2 \kappa} \cdot p_{m}q_{K} \log p_{m}q_{K})$ in our $<\chi_{S_{l,j,h}}, \hat{f}>$ calculations.  However, as long as the $<\chi_{S_{l,j,h}}, \hat{f}>$-measurement errors are small enough (i.e., of size $O(\delta \cdot B^{-p})$ in the $p$-compressible case) Theorem~\ref{thm:subfourier} and all related Section~\ref{sec:SigReconstruct} results and will still hold.  Using the proof of Theorems~\ref{thm:alg1} and \ref{thm:compress} along with some scratch work we can see that using $2 \kappa = O(\log \delta^{-1} + p)$ interpolation points per $f$-sample ensures all our $<\chi_{S_{l,j,h}}, \hat{f}>$-measurement errors are $O(\delta \cdot B^{-p})$.  We have the following result:

\begin{Theorem}
Let $\hat{\textbf{A}} = \Psi \textbf{A}$ be $p$-compressible.  Then, we may use Algorithms~\ref{alg:reconstruct} and \ref{alg:fmeasure} to return a $B$ term sparse representation, $\hat{\textbf{R}}^{\rm s}$, for $\hat{\textbf{A}}$ such that $\| \textbf{A} - \textbf{R} \|^{2}_{2} \leq \| \textbf{A} - \textbf{R}_{\rm opt} \|^{2}_{2} + \delta \| C^{\rm opt}_{B} \|^{2}_{2}$ using $\tilde{O} \left( B^{\frac{2p}{p-1}} \delta^{\frac{2}{1-p}} (\log \delta^{-1} + p)^{2} \right)$-time and $\tilde{O} \left( B^{\frac{2p}{p-1}} \delta^{\frac{2}{1-p}} (\log \delta^{-1} + p) \right)$-samples from \textbf{A}.
\label{thm:detDFT}
\end{Theorem}

Notice that Theorem~\ref{thm:detDFT} no longer guarantees an $\delta = O(\frac{1}{N})$-accurate $\tilde{O}(B^{2})$-time DFT algorithm for smooth data (i.e., \textbf{A}'s containing samples from a smooth function $f$).  This is because as $p \rightarrow \infty$ we require an increasingly large number of interpolation points per $f$-sample in order to guarantee our $<\chi_{S_{l,j,h}}, \hat{f}>$-measurements remain $O(\delta \cdot B^{-p})$-accurate.
However, for $\delta = O(\log^{-1} N)$, we can still consider smooth data \textbf{A} to be $O(\log N)$-compressible and so achieve a $\tilde{O}(B^{2})$-time DFT algorithm.  

\section{Conclusion}
\label{sec:conc}

Compressed Sensing (CS) methods provide algorithms for approximating the result of any large matrix multiplication as long as it is known in advance that the result will be sparse/compressible.  Hence, CS is potentially valuable for many numerical applications such as those involving multi-scale aspects \cite{SparseSpect,SparseSpectM}.  In this paper we used CS methods to develop the first known deterministic sub-linear time sparse Fourier transform algorithm.  In the process, we introduced a new deterministic Compressed Sensing algorithm along the lines of Cormode and Muthukrishnan (CM) \cite{CMDetCS1,CMDetCS2}.  
Our new deterministic CS algorithm improves on CM's algebraic compressibility results while simultaneously maintaining their results concerning exponential compressibility.

Compressed Sensing is closely related to hashing methods, combinatorial group testing, and many other algorithmic problems \cite{FirstDetCS,Crprecis}.  Thus, $K$-majority $k$-strongly selective collections of sets and Algorithm~\ref{alg:reconstruct} should help improve results concerning algebraically compressible (at each moment in time) stream hashing/heavy-hitter identification.  Further development of these/other algorithmic applications is left as future work.  It is also worthwhile to note that Monte Carlo Fourier results similar to those of \cite{AAFFT2} may be obtained by altering our measurement construction in Section~\ref{sec:Measurements}.  If we construct our $\mathcal{S}_{l}$ collections by using only a small subset of randomly chosen $q_{j}$'s we will still locate all sufficiently energetic entries of $\hat{\textbf{A}}$ with high probability.  The entries' coefficients can then be approximated by standard USFFT techniques \cite{AAFFT2,USFFT1,USFFT2,AUSFFTrev}.

\section{Acknowledgments}

We would like to thank Graham Cormode and S. Muthukrishnan for answering questions about their work.  We would also like to thank Martin Strauss, Anna Gilbert, Joel Lepak, and Hualong Feng for helpful discussions, advice, and comments.

\bibliographystyle{abbrv}
\bibliography{CSforSODA}

\appendix

\section{Proof of Theorem~\ref{thm:alg1}}

We begin by proving two lemmas.

\begin{Lemma}
IDENTIFICATION:  Lines 7 through 19 of Algorithm~\ref{alg:reconstruct} are guaranteed to recover all valid $\omega_{1}, \dots, \omega_{B}$ (i.e., all $\omega$ with $| \hat{\textbf{A}}(\omega) |_{2} \geq | \hat{\textbf{A}}(\omega_{B}) |_{2}$ - there may be $> B$ such entries) more then $\frac{2K}{3}$ times.  Hence, despite line 22, an entry for all such $\omega_{b}, 1 \leq b \leq B$, will be added to $\hat{\textbf{R}}^{\rm s}$ in line 26.
\label{lem:identification}
\end{Lemma}

\noindent \textit{Proof:}\\

Because of the construction of $\mathcal{S}_{0}$ (i.e., proof of Lemma~\ref{lem:S0strong}) we know that for each $b \in [1,B]$ there exist more then $\frac{2K}{3}$ subsets $S \in \mathcal{S}_{0}$ such that $S \cap \{ \omega_{b'}~|~b' \in [1,B'] \} = \{ \omega_{b}\}$.  Choose any $b \in [1,B]$.  Denote the $q$-primes that isolate $\omega_{b}$ from all of $\omega_{1}, \dots, \omega_{b-1}, \omega_{b+1}, \dots, \omega_{B'}$ by
$$q_{j_{1}}, q_{j_{2}}, \dots, q_{j_{K'}},~~\frac{2K}{3} < K' \leq K.$$
We next show that, for each $k' \in [1,K']$, we get $<\chi_{S_{0,j_{k'},\omega_{b} \textrm{ mod } q_{j_{k'}}}},\textbf{A}>$ as one of the $B' + 1$ largest magnitude $<\chi_{S_{0,j_{k'},\cdot}},\hat{\textbf{A}}>$-measurements identified in line 10.

Choose any $k' \in [1,K']$.  We know that
$$\frac{\epsilon}{2}~<~\frac{\epsilon}{\sqrt{2}}~<~|\hat{\textbf{A}}(\omega_{B})| - \sqrt{2} \sum^{N-1}_{b'=B'} | \hat{\textbf{A}}(\omega_{b'}) |~\leq~|\hat{\textbf{A}}(\omega_{b})| - \left| \sum_{b' \in [B',N),~\omega_{b'} \equiv \omega_{b}} \hat{\textbf{A}}(\omega_{b'}) \right|~\leq~\left| <\chi_{S_{0,j_{k'},\omega_{b} \textrm{ mod } q_{j_{k'}}}},\hat{\textbf{A}}> \right|.$$
We also know that the $(B'+1)^{\rm st}$ largest measurement L$^{2}$-magnitude must be $< \frac{\epsilon}{2}$.  Hence, we are guaranteed to execute lines 12-15 with an $r_{0,\cdot} = \omega_{b}$ mod $q_{j_{k'}}$.

Choose any $l \in [1,m]$ and set 
$$\Omega = \big\{ \omega_{b'}~\big|~b' \in [B',N),~\omega_{b'} \equiv \omega_{b} \textrm{ mod } q_{j_{k'}},~\omega_{b'} \nequiv \omega_{b} \textrm{ mod } q_{j_{k'}}p_{l}  \big\}.$$  
Using Lemma~\ref{lem:S1-m} we can see that line 13 inspects all the necessary residues of $\omega_{b}$ mod $p_{l}q_{j_{k'}}$.  To see that $t_{min}$ will be chosen correctly we note first that
$$\left| <\chi_{S_{0,j_{k'},\omega_{b} \textrm{ mod } q_{j_{k'}}}},\hat{\textbf{A}}> - <\chi_{S_{0,j_{k'},\omega_{b} \textrm{ mod } p_{l}q_{j_{k'}}}},\hat{\textbf{A}}> \right| = \left| \sum_{ \omega_{b'} \in \Omega} \hat{\textbf{A}}(\omega_{b'}) \right| ~\leq~\sqrt{2} \sum_{\omega_{b'} \in \Omega} | \hat{\textbf{A}}(\omega_{b'}) |.$$ 
Furthermore, setting $r_{0,\cdot} = \omega_{b}$ mod $q_{j_{k'}}$ and
$$\Omega ' = \big\{ \omega_{b'}~\big|~b' \in [B',N),~\omega_{b'} \equiv \omega_{b} \textrm{ mod } q_{j_{k'}},~\omega_{b'} \nequiv (r_{0,\cdot} + tq_{j_{k'}}) \textrm{ mod } q_{j_{k'}}p_{l} \textrm{ for some } t \textrm{ with } (r_{0,\cdot} + tq_{j_{k'}}) \nequiv \omega_{b} \textrm{ mod } q_{j_{k'}}p_{l} \big\},$$ 
we have
$$\sqrt{2} \sum_{\omega_{b'} \in \Omega} | \hat{\textbf{A}}(\omega_{b'}) | < \frac{\epsilon}{\sqrt{2}}~<~ |\hat{\textbf{A}}(\omega_{B})| - \sqrt{2} \sum^{N-1}_{b'=B'} |\hat{\textbf{A}}(\omega_{b'})|~\leq~|\hat{\textbf{A}}(\omega_{b})| - \left| \sum_{\omega_{b'} \in \Omega ' } \hat{\textbf{A}}(\omega_{b'}) \right|.$$
Finally we can see that 
$$ |\hat{\textbf{A}}(\omega_{b})| - \left| \sum_{\omega_{b'} \in \Omega '} \hat{\textbf{A}}(\omega_{b'}) \right| ~\leq~ \left| <\chi_{S_{0,j_{k'},\omega_{b} \textrm{ mod } q_{j_{k'}}}},\hat{\textbf{A}}> - <\chi_{S_{0,j_{k'},(r_{0,\cdot} + tq_{j_{k'}}) \nequiv \omega_{b} \textrm{ mod } p_{l}q_{j_{k'}}}},\hat{\textbf{A}}> \right|.$$
Hence, lines 13 and 14 will indeed select the correct residue for $\omega_{b}$ modulo $p_{l}$.  Therefore, line 16 will correctly reconstruct $\omega_{b}$ at least $K' > \frac{2K}{3}$ times.~~$\Box$ \\

\begin{Lemma}
ESTIMATION:  Every $(\omega, \tilde{\hat{{\rm A}}}_{\omega})$ stored in $\hat{\textbf{R}}^{\rm s}$ in line 27 is such that $|\hat{\textbf{A}}(\omega) - \tilde{\hat{{\rm A}}}_{\omega}|_{2} < \epsilon$.
\label{lem:estimation}
\end{Lemma}

\noindent \textit{Proof:} \\

Suppose that $(\omega, \tilde{\hat{{\rm A}}}_{\omega})$ is stored in $\hat{\textbf{R}}^{\rm s}$.  This only happens if $\hat{\textbf{A}}(\omega)$ has been estimated by $$<\chi_{S_{0,j,\omega \textrm{ mod } q_{j}}},\hat{\textbf{A}}> ~= \sum_{\tilde{\omega} \equiv \omega \textrm{ mod } q_{j}} \hat{\textbf{A}}(\tilde{\omega})$$ for more then $\frac{2K}{3}$ $q_{j}$-primes.  The only way that any such estimate can have $|\hat{\textbf{A}}(\omega) - <\chi_{S_{0,j,\omega \textrm{ mod } q_{j}}},\hat{\textbf{A}}>|_{1} ~\geq~ \frac{\epsilon}{\sqrt{2}}$ is if $\omega$ collides with one of $\omega_{1}, \dots, \omega_{B'}$ modulo $q_{j}$ (this is due to the definition of $B'$ in Equation~\ref{def:B'}).  By the proof of Lemma~\ref{lem:S0strong} we know this can happen at most $B' \lfloor \log_{B'} N \rfloor < \frac{K}{3}$ times.  Hence, more then half of the $\frac{2K}{3}$ estimates, $\tilde{\hat{{\rm A}}}_{\omega}'$, must be such that $|\hat{\textbf{A}}(\omega) - \tilde{\hat{{\rm A}}}_{\omega}'|_{1} < \frac{\epsilon}{\sqrt{2}}$.  It follows that taking medians as per lines 24 and 25 will result in the desired $\epsilon$-accurate estimate for $\hat{\textbf{A}}(\omega)$.
~~$\Box$ \\

We are now ready to prove Theorem~\ref{thm:alg1}.\\

\noindent \textbf{Theorem~\ref{thm:alg1}}~~\textit{Let $\hat{\textbf{R}}_{\rm opt}$ be a $B$-optimal Fourier representation for our input signal $\hat{\textbf{A}}$.  Then, the $B$ term representation $\hat{\textbf{R}}^{\rm s}$ returned from Algorithm~\ref{alg:reconstruct} is such that $\| \textbf{A} - \textbf{R} \|^{2}_{2} \leq \| \textbf{A} - \textbf{R}_{\rm opt} \|^{2}_{2} + \frac{6 B \cdot |\hat{\textbf{A}}(\omega_{B})|^{2}}{C}$.  Furthermore, Algorithm~\ref{alg:reconstruct}'s Identification and Estimation (lines 7 - 30) run time is $O(B'^{2} \log^{4} N )$.  The number of measurements used is $O( B'^{2} \log^{6} N )$}.\\

\noindent \textit{Proof:}\\

Choose any $b \in (0,B]$.  Using Lemmas~\ref{lem:identification} and~\ref{lem:estimation} we can see that only way some $\omega_{b} \notin \hat{\textbf{R}}^{\rm s}_{B}$ is if there exists some associated $b' \in (B,N)$ so that $\omega_{b'} \in \hat{\textbf{R}}^{\rm s}$ and
$$| \hat{\textbf{A}}(\omega_{B}) | + \epsilon \geq | \hat{\textbf{A}}(\omega_{b'}) | + \epsilon > | \tilde{\hat{{\rm A}}}_{\omega_{b'}}| \geq | \tilde{\hat{{\rm A}}}_{\omega_{b}} | > | \hat{\textbf{A}}(\omega_{b}) | - \epsilon \geq | \hat{\textbf{A}}(\omega_{B}) | - \epsilon.$$ 
In this case we'll have $2 \epsilon > | \hat{\textbf{A}}(\omega_{b}) | - | \hat{\textbf{A}}(\omega_{b'}) | \geq 0$ so that
\begin{equation}
| \hat{\textbf{A}}(\omega_{b'}) |^{2} + 4 \epsilon \left(  \epsilon + | \hat{\textbf{A}}(\omega_{B}) | \right) \geq | \hat{\textbf{A}}(\omega_{b'}) |^{2} + 4 \epsilon \left(  \epsilon + | \hat{\textbf{A}}(\omega_{b'}) | \right) > | \hat{\textbf{A}}(\omega_{b}) |^{2}.
\label{eqn:proofe}
\end{equation}

Now using Lemma~\ref{lem:estimation} we can see that
$$\| \hat{\textbf{A}} - \hat{\textbf{R}}\|^{2} = \sum_{(\omega, \cdot) \notin \hat{\textbf{R}}^{\rm s}} |\hat{\textbf{A}}(\omega)|^{2} + \sum_{(\omega, \tilde{\hat{{\rm A}}}_{\omega}) \in \hat{\textbf{R}}^{\rm s}} |\hat{\textbf{A}}(\omega) - \tilde{\hat{{\rm A}}}_{\omega} |^{2} < \sum_{(\omega, \cdot) \notin \hat{\textbf{R}}^{\rm s}} |\hat{\textbf{A}}(\omega)|^{2} + B \cdot \epsilon^{2}.$$
Furthermore, we have
$$B \cdot \epsilon^{2} + \sum_{(\omega, \cdot) \notin \hat{\textbf{R}}^{\rm s}} |\hat{\textbf{A}}(\omega)|^{2} = B \cdot \epsilon^{2} + \sum_{b \in (0,B],~\omega_{b} \notin \hat{\textbf{R}}^{\rm s}} |\hat{\textbf{A}}(\omega_{b})|^{2} + \sum_{b' \in (B,N),~\omega_{b'} \notin \hat{\textbf{R}}^{\rm s}} |\hat{\textbf{A}}(\omega_{b'})|^{2}.$$
Using observation \ref{eqn:proofe} above we can see that this last expression is bounded above by
$$B \cdot ( 5\epsilon^{2} + 4 \epsilon | \hat{\textbf{A}}(\omega_{B}) | ) + \sum_{b' \in [B,N),~\omega_{b'} \in \hat{\textbf{R}}^{\rm s}} |\hat{\textbf{A}}(\omega_{b'})|^{2} + \sum_{b' \in (B,N),~\omega_{b'} \notin \hat{\textbf{R}}^{\rm s}} |\hat{\textbf{A}}(\omega_{b'})|^{2} \leq \| \hat{\textbf{A}} - \hat{\textbf{R}}_{\rm opt} \|^{2}_{2} + B \cdot ( 5\epsilon^{2} + 4 \epsilon | \hat{\textbf{A}}(\omega_{B}) | ).$$
Substituting for $\epsilon$ (see Equation~\ref{eqn:epsilon}) gives us our result.  Mainly,

$$B \cdot ( 5\epsilon^{2} + 4 \epsilon | \hat{\textbf{A}}(\omega_{B}) | ) = \frac{B| \hat{\textbf{A}}(\omega_{B}) |^{2}}{C} \left( \frac{5}{2C} + 2 \sqrt{2} \right) < \frac{6 B| \hat{\textbf{A}}(\omega_{B}) |^{2}}{C}.$$

We next focus on run time.  Algorithm~\ref{alg:reconstruct}'s Identification (i.e., lines 7 through 19) run time is dominated by the $O(K B' m)$ executions of line 13.  And, each execution of line 13 takes time $O(p_{m})$.  Hence, given that $m = O(\log N)$, $p_{m} = O(\log N \cdot \log\log N)$, and $K = O(B' \log_{B'} N)$, we can see that Identification requires $O(B'^{2} \cdot \log_{B'} N \cdot \log^{2} N  \cdot \log \log N)$-time.

Continuing, Algorithm~\ref{alg:reconstruct}'s Estimation (i.e., lines 20 through 30) run time is ultimately determined by line 22's {\sc if}-statement.  Although line 22 is executed $O(K B') = O(B'^{2} \log_{B'} N)$ times, it can only evaluate to true $O(B')$ times.  Hence, each line 24/25 $O(B' \log_{B'} N \log B')$-time median operation will be evaluated at most $O(B')$ times.  The resulting Estimation runtime is therefore $O(B'^{2} \log_{B'} N \log B')$.

To bound the number of measurements we recall that: $(i)$ the number of measurements is $< m \cdot K \cdot p_{m}q_{K}$, $(ii)$ $m = O(\log N)$, $(iii)$ $p_{m} = O(\log N \cdot \log\log N)$, $(iv)$ $K = O( B' \log N)$, and $(v)$ $q_{K} = O(K \log K)$. Hence, the number of measurements is $O \left( K^{2} \log K \log^{2} N \log \log N \right)$.  Substituting for $K$ gives us the desired bound.
~~$\Box$ \\

\section{Proof of Theorem~\ref{thm:compress}}

\noindent \textbf{Theorem~\ref{thm:compress}}~~\textit{Let $\hat{\textbf{A}}$ be $p$-compressible.  Then, Algorithm~\ref{alg:reconstruct} can return a $B$ term sparse representation $\hat{\textbf{R}}^{\rm s}$ with $\| \textbf{A} - \textbf{R}\|^{2}_{2} \leq \| \textbf{A} - \textbf{R}_{\rm opt} \|^{2}_{2} + \delta \| C^{\rm opt}_{B} \|^{2}_{2}$ using $O \left( B^{\frac{2p}{p-1}} \delta^{\frac{2}{1-p}} \log^{4} N \right)$ total identification/estimation time and $O \left( B^{\frac{2p}{p-1}} \delta^{\frac{2}{1-p}} \log^{6} N \right)$ measurements.  If $\hat{\textbf{A}}$ decays exponentially, Algorithm~\ref{alg:reconstruct} can return a B term sparse representation, $\hat{\textbf{R}}^{\rm s}$, with $\| \textbf{A} - \textbf{R}\|^{2}_{2} \leq \| \textbf{A} - \textbf{R}_{\rm opt} \|^{2}_{2} + \delta \| C^{\rm opt}_{B} \|^{2}_{2}$ using both $\left( B^{2} + \log^{2} \delta^{\frac{-1}{\alpha}} \right) \cdot$ polylog($N$) measurements and identification/estimation time.}\\

\noindent \textit{Proof:} \\

We first deal with the algebraically compressible case.  We have to determine our Algorithm~\ref{alg:reconstruct}'s $B'$ and Theorem~\ref{thm:alg1}'s $C$ variables.  Moving toward that goal we note that
$$\frac{6 B \cdot |\hat{\textbf{A}}(\omega_{B})|^{2}}{C} = \frac{1}{C}O \left( B^{-2p+1} \right) = O \left( \frac{1}{C} \right) \| C^{\rm opt}_{B} \|^{2}_{2}.$$
After looking at Theorem~\ref{thm:alg1} we can see that we must use $C = O \left( \frac{1}{\delta} \right)$ and a $B'$ (see Equations~\ref{eqn:epsilon} and~\ref{def:B'}) so that
$$\sum^{N-1}_{b=B'} | \hat{\textbf{A}}(\omega_{b}) | = O(\delta \cdot |\hat{\textbf{A}}(\omega_{B})|) = O( \delta \cdot B^{-p} ).$$
Continuing,
$$\sum^{N-1}_{b = B'} |\hat{\textbf{A}}(\omega_{b})|_{2} = O \left( \int^{\infty}_{B'} b^{-p} db \right) = O( B'^{1-p} ).$$
Hence, we must use $B' = O \left( \delta^{\frac{1}{1-p}} B^{\frac{p}{p-1}} \right)$.  Applying Theorem~\ref{thm:alg1} gives us Algorithm~\ref{alg:reconstruct}'s runtime and number of required measurements.

We next deal with the exponentially compressible case.  Now, as before, we have to determine our Algorithm~\ref{alg:reconstruct}'s $B'$ and Theorem~\ref{thm:alg1}'s $C$ variables.  To do so we note that
$$\frac{6 B \cdot |\hat{\textbf{A}}(\omega_{B})|^{2}}{C} = \frac{B}{C} O \left( 4^{- \alpha B} \right) = O \left( \frac{B}{C} \right) \| C^{\rm opt}_{B} \|^{2}_{2}.$$
After looking at Theorem~\ref{thm:alg1} we can see that we must use $C = O \left( \frac{B}{\delta} \right)$ and a $B'$ so that
$$\sum^{N-1}_{b=B'} | \hat{\textbf{A}}(\omega_{b}) |_{2} = O \left( \frac{\delta \cdot |\hat{\textbf{A}}(\omega_{B})|}{B} \right) = O \left( \delta \cdot 2^{- \alpha B - \log B} \right).$$
Continuing,
$$\sum^{N-1}_{b = B'} |\hat{\textbf{A}}(\omega_{b})|_{2} = O \left( \sum^{\infty}_{b = B'} 2^{- \alpha b}\right) = O \left( 2^{- \alpha B'} \right).$$
Hence, we must use $B' = O \left( B + \log \left( \frac{B}{\delta} \right)^{\frac{1}{\alpha}} \right)$.  Applying Theorem~\ref{thm:alg1} gives us Algorithm~\ref{alg:reconstruct}'s runtime and number of required measurements.~~$\Box$ \\

\section{Proof of Theorem~\ref{thm:subfourier}}

\noindent \textbf{Theorem~\ref{thm:subfourier}}~~\textit{Let $f: [0,2 \pi] \rightarrow \mathbb{C}$ have $(i)$ an integrable $p^{\rm th}$ derivative, and $(ii)$ $f(0) = f(2 \pi), f'(0) = f'(2 \pi), \dots, f^{(p-2)}(0) = f^{(p-2)}(2 \pi)$ for some $p > 1$.  Furthermore, assume that $\hat{f}$'s $B' = O \left( B^{\frac{2p}{p-1}} \epsilon^{\frac{2}{1-p}} \right)$ largest magnitude frequencies all belong to $\left( - \big\lceil \frac{N}{2} \big\rceil, \big\lfloor \frac{N}{2} \big\rfloor \right]$.  Then, we may use Algorithm~\ref{alg:reconstruct} to return a $B$ term sparse Fourier representation, $\hat{\textbf{R}}^{\rm s}$, for $\hat{f}$ such that $\| \hat{f} - \hat{\textbf{R}} \|^{2}_{2} \leq \| \hat{f} - \hat{\textbf{R}}_{\rm opt} \|^{2}_{2} + \delta \| C^{\rm opt}_{B} \|^{2}_{2}$ using $O \left( B^{\frac{2p}{p-1}} \delta^{\frac{2}{1-p}} \log^{7} N \right)$-time and $O \left( B^{\frac{2p}{p-1}} \delta^{\frac{2}{1-p}} \log^{6} N \right)$-measurements from $f$.}\\

\noindent \textit{Proof:} \\

We note that Theorems~\ref{thm:alg1} and~\ref{thm:compress} still hold for $p$-compressible infinite signals/vectors $\hat{\textbf{A}}$ (i.e., signals with $\infty$-length).  For the purposes of proof we may consider $\hat{\textbf{A}}$ to be formed by any bijective mapping $g: \mathbb{N} \rightarrow \mathbb{Z}$ so that both
$$g \left( [0,N) \right) = \left( - \left\lceil \frac{N}{2} \right\rceil, \left\lfloor \frac{N}{2} \right\rfloor \right]$$
and
$$g(n) \equiv n \textrm{ mod } N, \textrm{ for all } n \in \mathbb{N},$$
are true.  We then set
$$\hat{\textbf{A}}(n) = \hat{f} \left( g(n) \right), \textrm{ for all } n \in \mathbb{N}.$$
In this case we note that $\hat{f}$'s $B'$ largest magnitude frequencies belonging to $\left( - \big\lceil \frac{N}{2} \big\rceil, \big\lfloor \frac{N}{2} \big\rfloor \right]$ implies our that $K$-majority $B'$-strongly separating collection of (infinite) subsets, $\mathcal{S}$, will still correctly isolate all the $B'$ most energetic frequency positions in $\hat{\textbf{A}}$.

Continuing, we may still consider our infinite length $p$-compressible, with $p > 1$, signal $\hat{\textbf{A}}$ (i.e., the Fourier coefficient series $\hat{f}$) to be sorted by magnitude for the purpose of identifying valid $\omega_{1}, \omega_{2}$, etc.. Furthermore, we may bound the (now infinite) sums of $\hat{\textbf{A}}$'s entries' magnitudes by the same integrals as above.  The proofs of the theorems/supporting lemmas will go through exactly as before if we consider all the finite sums in their proofs to be absolutely convergent infinite sums.  The only real difference from our work in Section~\ref{sec:SigReconstruct} is that we are computing our $<\chi_{S_{l,j,h}},\hat{\textbf{A}}>$-measurements differently.

Similar to Theorem~\ref{thm:alg1}, the required number of measurements from $f$ will be $O \left( B^{\frac{2p}{p-1}} \epsilon^{\frac{2}{1-p}} \log^{6} N \right)$.  This is exactly because for each $(q_{j}, p_{l})$-pair we compute all the measurements $$\left\{ <\chi_{S_{l,j,h}}, \hat{f}> ~\big|~ h \in [0, q_{j}p_{l}) \right\}$$
via one FFT requiring $q_{j}p_{l}$ samples.  Hence, the number of samples from $f$ we use is once again bounded above by $m \cdot K \cdot p_{m}q_{K}$.  Furthermore, each of the $m K$ FFT's will take $O( p_{m}q_{K} \log p_{m}q_{K} )$-time (despite the signal lengths' factorizations \cite{1968-bluestein,rabiner-schafer-rader}).  The result follows.
~~$\Box$ \\

\end{document}